\documentclass[aps,prl,10pt,twocolumn,floatfix,superscriptaddress]{revtex4-2}
\usepackage{graphicx}
\usepackage{bm}
\usepackage{amsmath}
\usepackage{mathrsfs}
\usepackage{txfonts}
\usepackage[english]{babel}
\usepackage{latexsym}
\usepackage{array}
\usepackage{multirow}
\usepackage{braket}
\usepackage{physics}
\usepackage{comment}
\usepackage{cancel}
\usepackage[dvipsnames]{xcolor}
\usepackage[colorlinks=true,citecolor=Blue,linkcolor=RubineRed,urlcolor=Blue]{hyperref}

\begin{document}

\title{Learning Generalized Statistical Mechanics with Matrix Product States}
\author{Pablo D\'iez-Valle}
\email{pablo.diez@csic.es}
\affiliation{%
Instituto de Física Fundamental IFF-CSIC, Calle Serrano 113b, Madrid 28006, Spain
}

\author{Fernando Mart\'inez-Garc\'ia}
\affiliation{%
Instituto de Física Fundamental IFF-CSIC, Calle Serrano 113b, Madrid 28006, Spain
}

\author{Juan Jos\'e Garc\'ia-Ripoll}
\affiliation{%
Instituto de Física Fundamental IFF-CSIC, Calle Serrano 113b, Madrid 28006, Spain
}

\author{Diego Porras}
\affiliation{%
Instituto de Física Fundamental IFF-CSIC, Calle Serrano 113b, Madrid 28006, Spain
}

\begin{abstract}

We introduce a variational algorithm based on Matrix Product States that is trained by minimizing a generalized free energy defined using Tsallis entropy instead of the standard Gibbs entropy. As a result, our model can generate the probability distributions associated with generalized statistical mechanics. The resulting model can be efficiently trained, since the resulting free energy and its gradient can be calculated exactly through tensor network contractions, as opposed to standard methods which require estimating the Gibbs entropy by sampling. 
We devise a variational annealing scheme by ramping up the inverse temperature, which allows us to train the model while avoiding getting trapped in local minima. We show the validity of our approach in Ising spin-glass problems by comparing it to exact numerical results and quasi-exact analytical approximations. Our work opens up new possibilities for studying generalized statistical physics and solving combinatorial optimization problems with tensor networks.

\end{abstract}

\maketitle

\textit{Introduction.---} Statistical mechanics describes complex phenomena involving large ensembles of particles, such as phase transitions. 
The usual subjects of study in statistical mechanics are physical systems characterized by the Boltzmann probability distribution. This motivated the development and study of algorithms for sampling these distributions, such as those based in Markov Chain Monte Carlo (MCMC)~\cite{kirkpatrick1983optimization, swendsen1986replica, hukushima2003population, machta2010population}. Additionally, Boltzmann machines~\cite{hinton1984boltzmann, ackley1985learning, hinton2012practical} are popular models for unsupervised machine learning that are trained to encode a target probability distribution into a Boltzmann distribution. A similar process can be implemented using Born machines~\cite{han2018unsupervised, liu2018differentiable, coyle2020born}, where these probability distributions are encoded in quantum wave functions using either quantum or quantum-inspired architectures. Therefore, statistical mechanics and the methods developed for its study are not only important from a fundamental perspective but also key for certain applications in combinatorial optimization and machine learning. While the study of Boltzmann distributions has been highly successful, the development of tools for studying alternative statistics has not been explored as much. 
Consequently, developing generative models that produce generalized statistics could be of great interest.
One example is Tsallis' statistics~\cite{tsallis1988possible}, which is motivated by the description of long-range interacting systems.
These generalized statistics are exciting not only because they reveal novel fundamental phenomena but also because of their potential applications in optimization~\cite{tsallis1996generalized}, machine learning~\cite{jose2020free}, economics~\cite{ludescher2011universal} and other fields~\cite{tsallis2022entropy}.
Moreover, generalized statistics may present advantages leading to more efficient numerical implementations than their Boltzmann counterparts, as we show in this paper.

In this Letter, we present a generative model based on a Matrix Product State (MPS) variational ansatz that approximates equilibrium probability distributions generated by Tsallis generalized statistical mechanics. The combination of Tsallis statistics and MPS methods has several advantages. Firstly, it leads to a variational method that permits an efficient calculation of the generalized entropy through tensor network contractions. 
This is not possible when considering the Gibbs entropy, which requires a sampling process for its estimation~\cite{liu2023tensor}, slowing the algorithm and introducing a statistical error in the estimated entropy value. Secondly, our scheme opens up a new computational method for the study of generalized statistical mechanics and novel many-body phenomena in non-extensive systems. Finally, it can also be the basis for new unsupervised machine learning schemes as well as combinatorial optimization methods.

\textit{Generalized entropy.---} Consider an Ising model with $N$ spins. Given a probability distribution parameterized by $\bm{\theta}$, $p_{\bm{\theta}}(\mathbf{s})$, for each spin configuration, $\mathbf{s}\in\{-1, +1\}^N$, the Tsallis entropy, $S_q(p_{\bm{\theta}})$, is defined as:
\begin{equation}
    S_q(p_{\bm{\theta}}) = \frac{1}{q-1}\left(1-\sum_\mathbf{s} p_{\bm{\theta}}(\mathbf{s})^q\right),
\end{equation}
which is characterized by the real parameter $q$~\cite{tsallis1988possible, tsallis2022entropy}.
This definition of entropy generalizes that of the Gibbs entropy, $S_{G}(p_{\bm{\theta}})\equiv -\sum_\mathbf{s} p_{\bm{\theta}}(\mathbf{s}) \log(p_{\bm{\theta}}(\mathbf{s}))$, which is recovered in the limit $q\rightarrow 1$. For this case, the probability distribution that minimizes the free energy at a given inverse temperature $\beta=1/T$ is the Boltzmann distribution $p(\mathbf{s})=\exp\left[-\beta E(\mathbf{s})\right]/Z$, where $E(\mathbf{s})$ is the energy of a spin configuration and $Z$ is the partition function. Following this approach, it is possible to train a variational model to obtain approximate thermal distributions. A basic example of this approach is the mean-field method~\cite{opper2001naive}, which assumes independent probabilities for each spin, $p_{\bm{\theta}}(\mathbf{s})=\prod^N_{i=1} p_{i,\bm{\theta}}(s_i)$, where $p_{i,\bm{\theta}}(s_i)$ is the marginal probability for the value of the $i$-th spin. More advanced approaches capable of capturing spin correlations have been shown using neural networks~\cite{wu2019solving, pan2021solving} and tensor networks~\cite{liu2023tensor}.

\begin{figure}[t]
\includegraphics[width=1.0\linewidth]{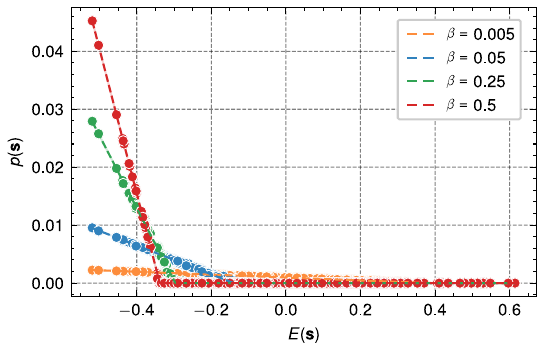}
\caption{Examples of the generalized thermal distribution in Eq.~\eqref{eq_generalized_thermal_distribution} for different inverse temperatures and $N=10$. The discontinuous lines represent the exact value of the probabilities as a function of the energy. We also show the results of our proposed variational model for a bond dimension $\chi=12$, represented as circles.}
\label{fig_bond_distribution}
\end{figure}

In this Letter, we will consider the case with $q=2$. Using this entropy, the variational free energy is given by:
\begin{equation}
    F^{\beta}_{q=2}(p_{\bm{\theta}})=\sum_\mathbf{s} E(\mathbf{s}) p_{\bm{\theta}}(\mathbf{s}) - \frac{1}{\beta}\left(1-\sum_\mathbf{s} p_{\bm{\theta}}(\mathbf{s})^2\right)\,,
    \label{eq_q2free_energy}
\end{equation}
It can be proven (see the original work~\cite{tsallis1988possible}, and Ref.~\cite{jose2020free} for a pedagogical derivation in a machine learning context) that the probability distribution that minimizes this free energy is given by:
\begin{equation}
\label{eq_generalized_thermal_distribution}
    p_{q=2}(\mathbf{s})=\max\left[-\frac{\beta}{2} E(\mathbf{s})+\tau, 0\right],
\end{equation}
where $\tau$ is a normalization factor with a similar role as $Z$ for the Boltzmann distribution. As a result of using this entropy, the probability of the different configurations behaves linearly with respect to their corresponding energies, with a slope given by $-\beta/2$ (see Fig.~\ref{fig_bond_distribution}). An interesting property is that, for increasing values of $\beta$, high-energy configurations could have a zero probability associated.

\textit{Tsallis statistics analysis.---} Firstly, we study the behaviour of instances of Ising models with energy $E(\mathbf{s})=-1/2\sum_{i,j}J_{ij} s_i s_j$ under Tsallis statistics. The interaction matrix $J$ is defined by a regular graph model of degree $d=6$ with random couplings, meaning that each spin variable shares six bonds with exactly six different spins. This architecture ensures the nondeterministic polynomial time hardness (NP-hardness) of the model simulation~\cite{barahona1982computational,jauma2024exploring}. For this regular graph case, we choose the value of each bond $J_{ij}$ connecting the $i$-th spin with the $j$-th spin to be generated from a normal distribution with zero mean and 1/$(Nd)$ variance so that the minimum energy is approximately constant (up to finite size effects).

\begin{figure}[t]
\includegraphics[width=1.0\linewidth]{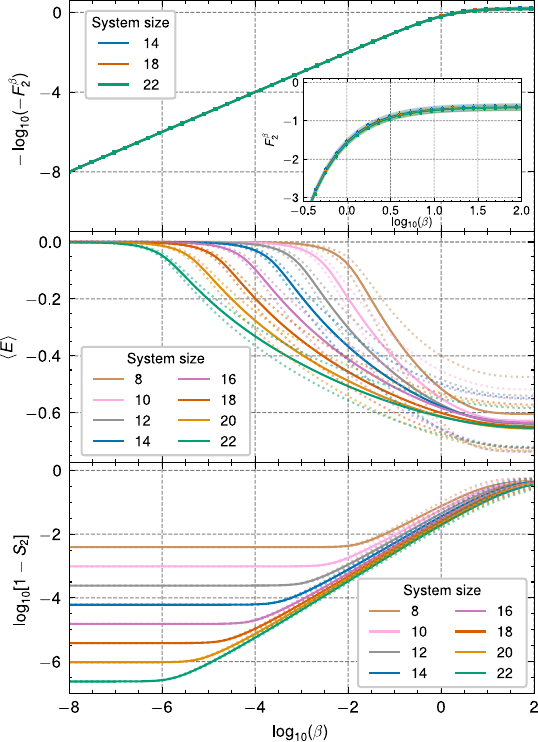}
\caption{Exact numerical results of the behaviour of different properties under Tsallis statistics as a function of $\beta$ for different system sizes averaged over $500$ instances of 6-regular problems for each size. Top: Generalized free energy estimated from the analytical approximation (squares) and exact results (continuous lines). Middle: Mean energy.  Bottom: Generalized entropy. All plots include the standard deviation of the results in the form of a shaded area (top) or dotted lines (middle and bottom).}
\label{fig_TsallisStatistic}
\end{figure}

We examine this model for small system sizes that allow for the exact numerical calculation of all its statistical properties. Additionally, we compare these results with an analytical approximation obtained from considering that the energy density of states is given by a normal distribution with a variance that can be easily estimated through sampling at $\beta=0$. We further improve this approximation by ignoring the terms of this distribution that correspond to energies smaller than the minimum energy (see~\cite{suppl}). The results are shown in Fig.~\ref{fig_TsallisStatistic}. 
We observe that the evolution of the generalized free energy with $\beta$ collapses with the system size in a common trajectory. Moreover, the analytical approximation is in close agreement with the exact results. As for the mean energy and generalized entropy, both values are constant until reaching a value of $\beta$ that approximates the maximum entropy of the system $1-S_2 = 2^{-N}$. Below this value, the system remains in the zero mean energy state, in which all spin configurations are equally probable. Beyond this threshold, progressive cooling produces lower mean energies and entropies, guiding the distribution towards the ground energy state of the system. The model studied has a $\mathbb{Z}_2$ symmetry, so the entropy asymptotically approaches $1-S_2 = 0.5$ as $\beta$ goes to infinity. Having understood these properties of the Tsallis statistics for our model, we emulate such dynamics for larger systems that are not amenable to exact computations. To this end, we introduce a variational MPS algorithm capable of generating the Tsallis statistic with high accuracy.

\textit{Matrix Product States algorithm.---} Tensor networks (TNs)~\cite{orus2014practical} are a mathematical framework used for the representation and manipulation of large multidimensional arrays of numbers, allowing for simplified computations. In the field of quantum mechanics, TNs are a popular and powerful tool to simulate the physical properties of correlated quantum many-body systems, but their study has also been extended to other fields such as machine learning~\cite{han2018unsupervised, liu2019machine, cheng2019tree, reyes2021multi}. The Matrix Product State (MPS)~\cite{perez2006matrix}, also known as Tensor Train (TT)~\cite{oseledets2011tensor}, is one of the most studied TN architectures for which several efficient algorithms have been designed. We use an MPS to represent a parameterized wave function of $N$ qubits by using a set of three-dimensional tensors $\theta_{\alpha_i\alpha_j}^{s}$ as:
\begin{equation}
    \ket{\Psi_{\boldsymbol{\theta}}} = \sum_{{\bf s} = {\{0,1\}}^{N}}\sum_{\alpha_1, \dots, \alpha_N =1}^{\chi} \left[\prod_{n=1}^{N} \theta^{[n]s_n}_{\alpha_{n}\alpha_{n+1}} \ket{\mathbf{s}}\right]\, ,
    \label{eq_MPS}
\end{equation}
where $\ket{\mathbf{s}}=\ket{s_1,s_2,...,s_{N}}$ is a quantum state representing a spin configuration, $\bm{\theta}=(\boldsymbol{\theta}^{[1]},..., \boldsymbol{\theta}^{[N]})$ is the set of tensors with a total of $2(N-2)\chi^2 + 4\chi$ parameters and $\chi$ is known as the bond dimension or tensor-train rank, which controls the expressivity of the MPS ansatz. The parameterized probability of sampling a spin configuration $\ket{\textbf{s}}$ is:
\begin{equation}
p_{\boldsymbol{\theta}}(\mathbf{s}) = \dfrac{|\braket{\mathbf{s}}{\Psi_{\boldsymbol{\theta}}}|^2}{|\Psi_{\boldsymbol{\theta}}|^2} = \frac{1}{|\Psi_{\boldsymbol{\theta}}|^2}\sum_{\alpha_1,\dots,\alpha_N=1}^{\chi} \left[\prod_{n=1}^{N} \theta^{[n]s_n}_{\alpha_{n}\alpha_{n+1}}\theta^{\dagger[n]s_n}_{\alpha_{n}\alpha_{n+1}}\right] \,,    
\end{equation}
where we have introduced the normalization factor $|\Psi_{\boldsymbol{\theta}}|^2$, which ensures that $\sum_{\mathbf{s}}p_{\boldsymbol{\theta}}(\mathbf{s})=1$.

\begin{figure}[t]
\includegraphics[width=0.95\linewidth]{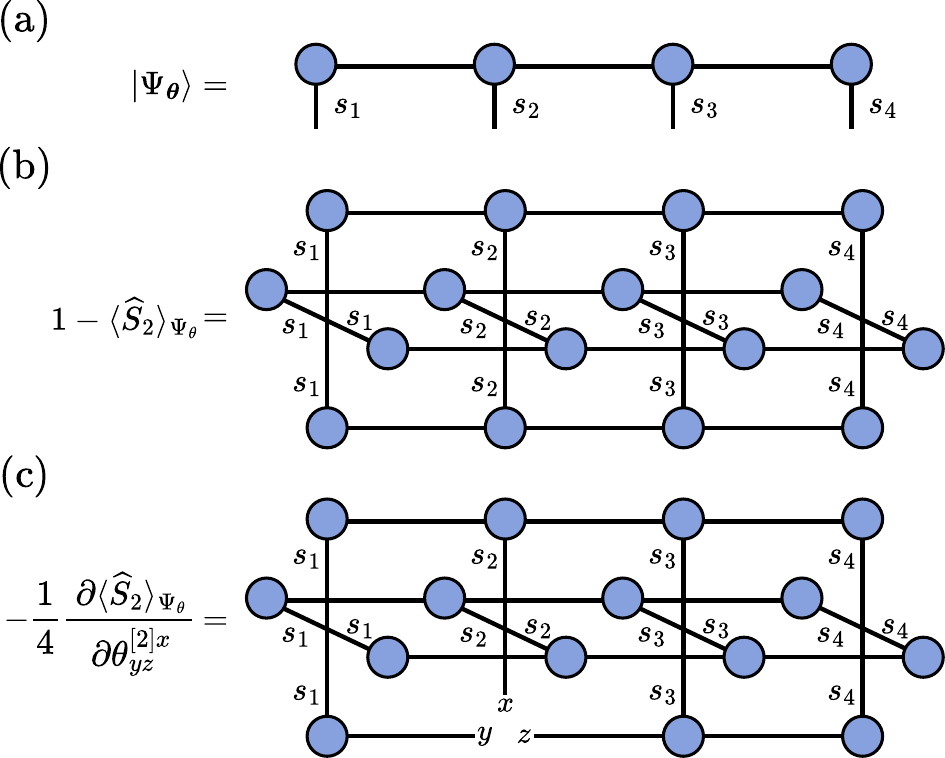}
\caption{(a) Schematic representation of the tensor network used to implement our variational model for a small size $N=4$. (b) Representation of the normalized tensor network contraction ($|\Psi_{\boldsymbol{\theta}}|^2 =1$) performed to obtain the Tsallis entropy with $q=2$ without having to sample configurations. (c) Representation of how to calculate the gradient of $S_2$ with respect to, in this example, the second site. In~\cite{suppl} we expand the case when the tensor network is not normalized.}
\label{fig:tensor_network}
\end{figure}

This MPS is trained by optimizing the set of parameters $\bm{\theta}$ to minimize the generalized variational free energy in Eq.~\eqref{eq_q2free_energy} with inverse temperature $\beta$. The training is based on iteratively optimizing individual MPS tensors (called \textit{sites} in the following), similar to the procedure followed by the Density Matrix Renormalization Group (DMRG) algorithm~\cite{white1992density, schollwock2005density}. When the parameters of site $n$ are optimized, the rest of the MPS sites are untouched. The minimization corresponds to:
\begin{equation}
    \min_{\boldsymbol{\theta}^{[n]}}  F_2^{\beta}(\boldsymbol{\theta})\,\quad \,\textnormal{with}\,\quad \, F_2^{\beta}(\boldsymbol{\theta})=\langle \hat{E} \rangle_{\Psi_{\boldsymbol{\theta}}} - \beta^{-1} \langle \hat{S}_2 \rangle_{\Psi_{\boldsymbol{\theta}}} \,,
\end{equation}
where $\langle \hat{O} \rangle_{\Psi_{\boldsymbol{\theta}}} \equiv \bra{\Psi_{\boldsymbol{\theta}}}\hat{O} \ket{\Psi_{\boldsymbol{\theta}}}/|\Psi_{\boldsymbol{\theta}}|^2$ with $\hat{O}\in\{\hat{E},\hat{S}_2\}$, $\hat{E} = \sum_{\mathbf{s}}E(\mathbf{s})\ketbra{\mathbf{s}}{\mathbf{s}}$, and $\hat{S}_2 = \mathbb{I} - \sum_{\mathbf{s}}p_{\boldsymbol{\theta}}(\mathbf{s}) \ketbra{\mathbf{s}}{\mathbf{s}}$. 

An important advantage of TNs is that they are an ideal framework for minimizing the generalized free energy since its gradient, $\nabla_{\boldsymbol{\theta}^{[n]}}F = \nabla_{\boldsymbol{\theta}^{[n]}}\langle \hat{E} \rangle_{\Psi_{\boldsymbol{\theta}}} - \beta^{-1} \nabla_{\boldsymbol{\theta}^{[n]}} \langle \hat{S}_2 \rangle_{\Psi_{\boldsymbol{\theta}}}$, can be computed exactly by performing index contractions. As shown in Fig.~\ref{fig:tensor_network} and~\cite{suppl}, the computation of the Tsallis entropy $\langle \hat{S}_2 \rangle_{\Psi_{\boldsymbol{\theta}}}$ and its derivative $\nabla \langle \hat{S}_2 \rangle_{\Psi_{\boldsymbol{\theta}}}$ scales linearly with the system size $N$ as $\mathcal{O}(N\chi^5)$ using an MPS structure.  In contrast, the Gibbs entropy $S_{G}$ cannot be calculated by similar schemes, and estimating the standard free energy requires sampling the distribution generated by the tensor network~\cite{liu2023tensor}. Although samples can be drawn from the probability distribution generated by the MPS efficiently, as explained in Ref.~\cite{han2018unsupervised}, iteratively repeating the sampling during the variational algorithm can introduce significant overhead. Additionally, considering a finite number of samples introduces an error in the entropy estimate, which damages the training process.

We start the optimization process at site $n=1$ and repeat it until site $N$ (left-to-right direction). At that point, the same iterations continue, changing the update to a right-to-left direction. We refer to each complete round from $1$ to $N$ or vice versa as a sweep. Training continues until a maximum number of sweeps or a convergence criterion (such as the variation of $F^{\beta}_2$ between sweeps) is satisfied. In the numerical results shown in this Letter, a variation of $F^{\beta}_2$ less than $10^{-4}$ is used as a convergence criterion. In practice, using the canonical form~\cite{verstraete04prl,catarina2023density} bounded by a maximum bond dimension $\chi$ with respect to site $n$ enhances the stability of the algorithm. Therefore, we update and normalize the MPS in canonical form at each site iteration. Note that all the involved contractions can be calculated at the beginning of the training and reused during the process. Thus, at each iteration, it is only necessary to perform the computations involving the MPS site $n$.

\begin{figure}[t]
\includegraphics[width=1.0\linewidth]{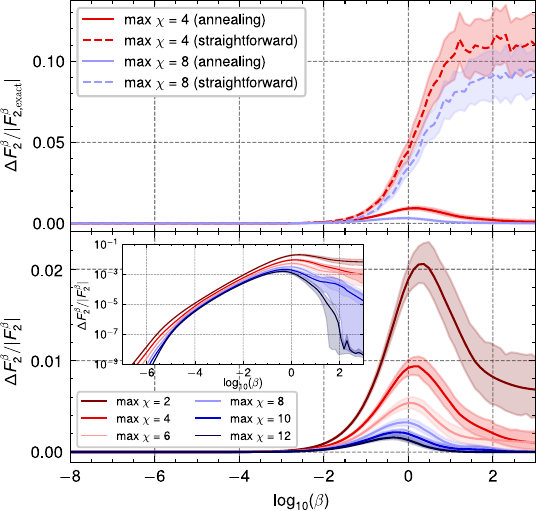}
\caption{Results of the free energy relative error for models of size $N=22$ and averaged over 100 samples. Top: Comparison of the behaviour of the relative error for two values of the maximum bond dimension $\chi$ using an annealing process (continuous lines) and without annealing (dashed lines).
Bottom: Behaviour of the relative error as a function of $\beta$ for different values of the maximum bond dimension for the annealed training of the variational model. Both plots include the 95\% confidence interval in the average estimation in the form of a shaded area.}
\label{fig_bond_FEerror}
\end{figure}

The above training method may be applied to a tensor network with initial random values of its parameters to learn the probability distribution for a target inverse temperature $\beta_f$. However, following this approach is not optimal since the variational model might easily get trapped in a local minimum, especially for high $\beta_f$. To avoid such an issue, we propose an annealing approach that starts from a small inverse temperature value, $\beta_0$, and gradually increases it while training the model. For each inverse temperature, the model is trained until reaching the convergence criterion defined before. When this happens, the inverse temperature is increased to the next value. This is repeated up to the desired final temperature $\beta_f$. As shown in Fig.~\ref{fig_bond_FEerror}, doing an annealed training results in a considerable improvement of the algorithm for high values of the target value $\beta_f$.

We implement this algorithm using the \textit{SeeMPS}~\cite{Seemps2024}, \textit{opt-einsum}~\cite{g_a_smith_opt_einsum_2018}, and \textit{SciPy}~\cite{virtanen_scipy_2020} Python libraries. The variational parameters are optimized via the gradient-based Limited-memory Broyden–Fletcher–Goldfarb–Shanno (L-BFGS) method~\cite{byrd_limited_1995}. The code is available at Ref.~\cite{Learning-Generalized-Statistical-Mechanics-with-MPS_repo}.

\begin{figure}[t]
\includegraphics[width=1\linewidth]{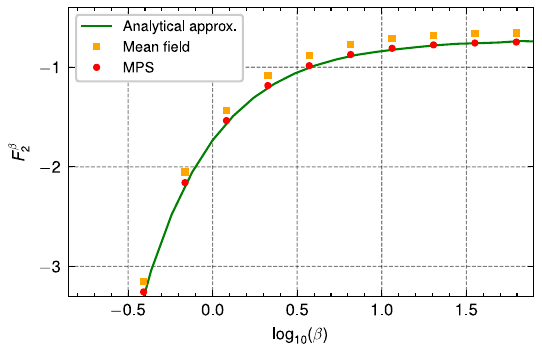}
\caption{Generalized free energy as a function of $\beta$ obtained from the MPS model with maximum bond dimension $\chi=6$ for a model of size $N=46$ compared to the results of the mean-field model and the analytical approximation. The results highlight the improvement achieved by our model over the mean-field approximation.}
\label{fig:MPS_results}
\end{figure}

\textit{Numerical results.---} We apply our proposed variational MPS algorithm to a problem of size $N=46$. In Fig.~\ref{fig:MPS_results} we compare the results obtained from the MPS algorithm to the analytical approximation, as well as the result obtained by a mean-field model obtained by setting $\chi=1$ in our model. We observe that the MPS model, which can capture spin correlations, improves over the results obtained from the mean-field model. Moreover, these results get close to the value estimated from the analytical approximation. It is important to note that the analytical approximation is closer to the exact value of the generalized free energy for both small and high values of $\beta$, with intermediate regions behaving more as a lower bound of the exact value~\cite{suppl}. This is a reason that causes the slight deviations of the MPS results from the analytical approximation behaviour for this region.

\begin{figure}[t]
\includegraphics[width=1.0\linewidth]{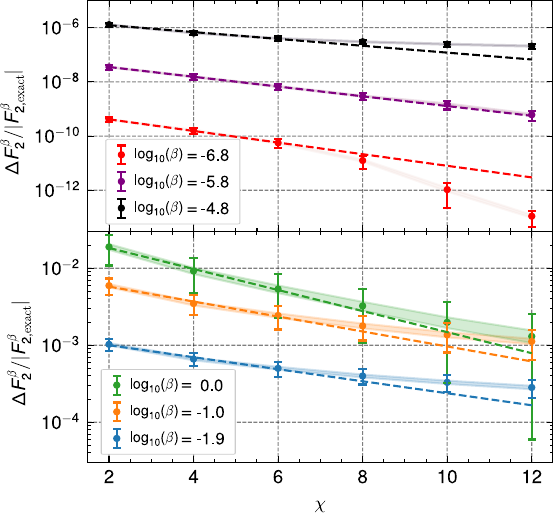}
\caption{Relative error of the generalized free energy as a function of the maximum bond dimension of the MPS for different values of $\beta$. The discontinuous lines are obtained from an exponential fit for the first three bond dimensions considered, $\chi=2,4,6$. These values are obtained from averaging the results of $100$ samples with $N=22$ spins. The shaded area corresponds to the 95\% confidence interval in the average estimation, while the bars correspond to the standard deviation of the average results.}
\label{fig_FEerror_bond_scaling}
\end{figure}

We also study the quality of the variational generalized free energy behaviour obtained from our model as a function of the bond dimension. We consider Ising models with $N=22$ spins to calculate the exact generalized free energy at each temperature step numerically. We obtain the variational free energy and the exact free energy and calculate the relative error during the annealing process for different values of the bond dimension, $\chi$ (see Fig.~\ref{fig_bond_FEerror}). For small values of $\beta$, the variational model generates a probability distribution with a generalized free energy that is in close agreement with that of the ideal distribution. This is expected since these models at high temperatures have small spin correlations, resulting in an easy thermalization. As $\beta$ increases, the relative error increases until it reaches a maximum. For these values of $\beta$, the model has more problems in approximating the ideal distribution. However, increasing the bond dimension of the model alleviates this problem. Finally, as $\beta$ reaches high enough values, only a few states should be associated with a non-zero probability, resulting in an easier-to-generate probability distribution. This results in a decrease in the relative error. However, as a consequence of the difficulties of generating the ideal distribution at intermediate temperatures, a residual error can appear at the end. This error is associated with the hardness of the optimization in the high $\beta$ regime due to the existence of local minima. Nevertheless, this residual error is again suppressed as the bond dimension is increased. We also show the behaviour of the relative error when the bond dimension is increased for several values of $\beta$ in Fig.~\ref{fig_FEerror_bond_scaling}. When $\beta$ approaches 1, the error discussed above, and associated with the local minima rather than the expressivity of the MPS, leads to a significant increase in the standard deviation of the results. 

\textit{Conclusions and outlook.---} In this Letter, we have introduced a tensor network variational model as a tool for generating generalized thermal distributions based on Tsallis entropy. While other variational methods based on neural networks~\cite{wu2019solving, pan2021solving} and tensor networks~\cite{liu2023tensor} have been studied previously, they consider the Gibbs entropy, whose estimation requires a sampling process at each training step. Apart from being computationally costly, this sampling requirement introduces a statistical error due to the finite size of the considered samples. In comparison, we have shown that Tsallis entropy can be estimated through tensor network contractions, resulting in an efficient training of the tensor network model.

Using an MPS, we have shown that our variational model minimizes the generalized free energy for different temperature values, thus approximating the correct generalized thermal distribution. This model yields better results than a mean-field model and the generalized free energies obtained agree with those derived from an analytical approximation of the ideal values, showing that the ability of the tensor network to capture spin correlations improves the quality of the results. Moreover, we have demonstrated how the accuracy of the results is improved when enhancing the capacity of the MPS by increasing its bond dimension.

Based on these positive results, it would be interesting to use this variational approach for studying generalized statistics in systems of interest in statistical mechanics, such as spin glasses~\cite{mezard1987spin, edwards1975theory}, which have usually been analysed considering the Boltzmann distribution. Moreover, we note that while we have focused on the study of spin systems, our proposed model can be easily generalized to other cases such as the Potts model~\cite{potts1952some, wu1982potts}. Another interesting route would be the study of this algorithm for solving combinatorial optimization problems. Finally, we note that further improvements could be achieved by the implementation of other tensor network models such as autoregressive architectures~\cite{liu2023tensor}, which we leave open for future work.

The code and data to reproduce the results of this Letter are available at Ref.~\cite{Learning-Generalized-Statistical-Mechanics-with-MPS_repo}.

\textit{Acknowledgments.---}  This project has been supported by the Spanish CDTI through Misiones Ciencia e Innovación Program (CUCO) under Grant MIG-20211005; Spanish Projects No. PID2021-127968NB-I00 and No. PDC2022-133486-I00, funded by MCIN/AEI/10.13039/501100011033 and by the European Union “NextGenerationEU”/PRTR”1; and 
CSIC Interdisciplinary Thematic Platform (PTI) Quantum Technologies (PTI-QTEP). We acknowledge both the Scientific Computing Area (AIC), SGAI-CSIC, for their assistance while using the DRAGO Supercomputer and Centro de Supercomputación de Galicia (CESGA) who provided access to the supercomputer FinisTerrae for performing the simulations. We also acknowledge the BBVA Quantum team for useful discussions.

\bibliography{sample}

\clearpage
\onecolumngrid

\begin{center}
  {\large\textbf{Supplementary material to ``Learning Generalized Statistical Mechanics with Matrix Product States''}}
\end{center}

\begin{center}
Pablo D\'iez-Valle$^1$, Fernando Mart\'inez-Garc\'ia$^1$, Juan Jos\'e Garc\'ia-Ripoll$^1$, and Diego Porras$^1$

  \textit{$^1$Instituto de F\'{\i}sica Fundamental IFF-CSIC, Calle Serrano 113b, Madrid 28006, Spain}\\
\end{center}
\vspace{0.5cm}

In this document, we extend and complete the results of the work presented in "Learning Generalized Statistical Mechanics with Matrix Product States". In Appendix I, we develop the analytical approximation to Tsallis statistics used as a benchmark in the main results of the manuscript. Furthermore, we provide a more detailed explanation of the Matrix Product State algorithm, as well as an analytical and numerical study of the cost scaling of the algorithm in Appendix II.
\\

\section{I. Analytical approximation to the Tsallis statistics}
\label{appendix_Tsallisanalyticalapprox}

The exact numerical calculation of universal Ising model observables evolved under Tsallis statistics at certain values of $\beta$ is infeasible for large systems. Therefore, we construct an analytical approximation of the model to provide an intuition of the correct trend of our tensor network algorithm results. The accuracy of the approximation is compared to brute-force computations on small scales.

\begin{figure}
\includegraphics[width=1\linewidth]{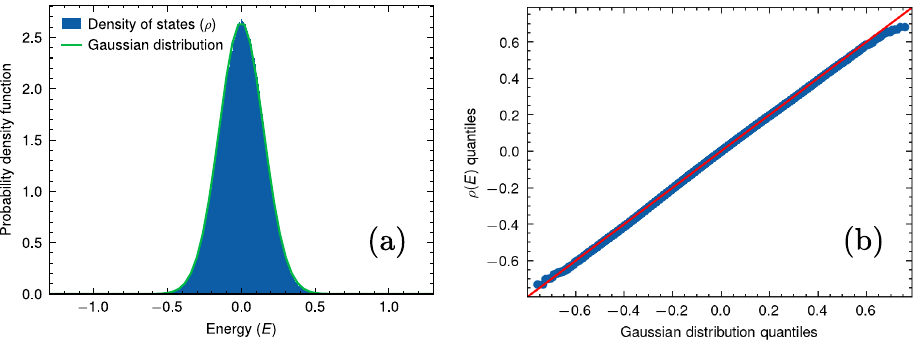}
\caption{Comparison between a Gaussian distribution and the density of states $\rho$ of a single instance of an Ising model defined by a regular graph of degree 6 and 22 spin variables. (a) Probability density function of both distributions. (b) Quantile-quantile plot to compare both distributions. The quantiles are shown by the blue dots, while the red line is the line of slope one that would match the results if the two distributions were equal. We can see how the density of states is well described over most of the energy spectrum, except for some deviations at the extremes (Color online).}
\label{Fig:denstates_gaussian}
\end{figure}

Let us assume that the density of states, $\rho$, of an Ising model is well described by a Gaussian distribution centred around zero:
\begin{equation}
    \rho(E) = \frac{1}{\sigma\sqrt{2\pi}}e^{\frac{- E^2}{2\sigma^2}}\;,
    \label{eq_appendix_gaussiandensitystates}
\end{equation}
where $\sigma$ is the standard deviation of the energies $E$. This definition is accurate for most of the energy spectrum but fails to describe the extremes (higher and lower energy states), as shown in Figure~\ref{Fig:denstates_gaussian}. This deviation causes small inaccuracies in the approximation for some regimes, as explained below. The equilibrium probability distribution that minimizes the generalized free energy with $q=2$ is:
\begin{equation}
    p_{q=2}(E)=\max\left[-\frac{\beta}{2} E+\tau, 0\right],
\end{equation}
where $\tau$ is a normalization factor whose straightforward calculation requires an exponential cost with the system size. However, with the density of states in Eq.~\eqref{eq_appendix_gaussiandensitystates} at hand, the normalization condition allows us to extract the value of $\tau$ by integrating over the entire spectrum and numerically solving the resulting equation: 
\begin{equation}
    2^N\int^{2\tau/\beta}_{-\infty} \rho(E) p(E) dE = e^{-\frac{2\tau^2}{\beta^2\sigma^2}}\frac{\beta\sigma}{2\sqrt{2\pi}} + \frac{1}{2\tau}\left(1+\erf{\left[\frac{\sqrt{2}\tau}{\beta\sigma}\right]}\right) = 1 \;,
\end{equation}
where $N$ is the number of variables in the system, and $\erf(x)$ is the error function. The approximation is further improved if we take into account the minimum energy of the system, $E_{min}$, which for the purpose of this manuscript is estimated using a carefully calibrated simulated annealing algorithm:
\begin{equation}
   2^N \int^{2\tau/\beta}_{E_{min}} \rho(E) p(E) dE = \left(e^{-\frac{2\tau^2}{\beta^2\sigma^2}}-e^{-\frac{E_{min}^2}{2\sigma^2}}\right)\frac{\beta\sigma}{2\sqrt{2\pi}} + \frac{1}{2\tau}\left(\erf{\left[\frac{\sqrt{2}\tau}{\beta\sigma}\right]}-\erf{\left[\frac{E_{min}}{\sqrt{2}\sigma}\right]}\right) = 1 \;,
\end{equation}

Once we have calculated the normalization factor of the probability distribution $\tau$, we can similarly approximate several other quantities that characterize the system such as the average energy,
\begin{equation}
    \langle E \rangle = 2^{N} \int^{2\tau/\beta}_{-E_{min}} \rho(E) p(E) E dE = \frac{\sigma}{4}\left(e^{-\frac{E_{min}^2}{2\sigma^2}} \sqrt{\frac{2}{\pi}}\left(2\tau-\beta E_{min}\right) + \beta\sigma\left(\erf{\left[\frac{E_{min}}{\sqrt{2}\sigma}\right]}-\erf{\left[\frac{\sqrt{2}\tau}{\beta\sigma}\right]}\right)\right) \;,
\end{equation}
or the Tsallis entropy,
\begin{equation}
    1 - S_2 = 2^{N} \int^{2\tau/\beta}_{-E_{min}} \rho(E) p^2(E) dE = \frac{\beta\sigma}{4\sqrt{2\pi}} \left(e^{-\frac{E_{min}^2}{2\sigma^2}} \left(\beta E_{min}-4\tau\right) + 2\tau e^{-\frac{2\tau^2}{\beta^2\sigma^2}} \right) + \frac{1}{8}\left(\beta^2\sigma^2 + 4\tau^2\right)\left(\erf{\left[\frac{\sqrt{2}\tau}{\beta\sigma}\right] } -\erf{\left[\frac{E_{min}}{\sqrt{2}\sigma}\right]} \right) \;.
\end{equation}

\begin{figure}
\includegraphics[width=1\linewidth]{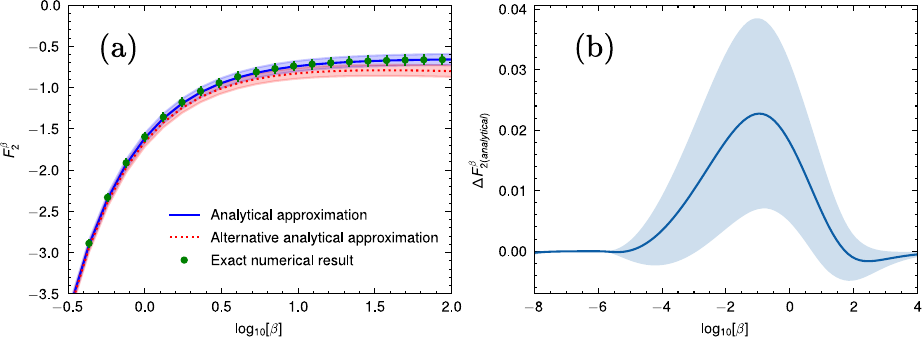}
\caption{Analytical approximation accuracy in the calculation of the generalized free energy $F^{\beta}_2$. We show the average results for 500 instances of 22-spin Ising models defined by regular graphs of degree 6. All results show the standard deviation as shaded areas or bars. (a) Comparison of the result of the analytical approximation taking into account the minimum energy (solid line), without the minimum energy (dashed line), and the exact numerically calculated result (dots). (b) Quantitative error of the analytical approximation in the estimate of the generalized free energy, $\Delta F^{\beta}_{2(analytical)} = F^{\beta}_{2(exact)} - F^{\beta}_{2(analytical)}$ (Color online).}
\label{Fig:analytical_approacherror}
\end{figure}

In Figure~\ref{Fig:analytical_approacherror} we show a qualitative and quantitative estimate of the error made by the above analytical approximation in the calculation of the generalized free energy $F^{\beta}_2$. As can be seen, the approximation provides good accuracy in the estimation of the generalized free energy. Furthermore, the inclusion of the minimum energy allows us to improve the results in the high $\beta$ regime. The analytical approximation is especially accurate for low and high $\beta$ stages, exhibiting slightly larger deviations in the $\beta \in [-2,0]$ regime where it acts as a lower bound on the exact result for most instances. 

\section{II. Computational cost of the MPS algorithm}
\label{appendix_scalings}

Given a Hamiltonian $\hat{E}$ and an MPS $\ket{\Psi_{\boldsymbol{\theta}}}$ with squared norm $|\Psi_{\theta}|^2\equiv|\braket{\Psi_{\boldsymbol{\theta}}}{\Psi_{\boldsymbol{\theta}}}|^2$, the generalized free energy with $q=2$ in Eq.~\eqref{eq_q2free_energy} can be expressed as:
\begin{equation}
    F_2^{\beta}(\boldsymbol{\theta}) = \langle \hat{E} \rangle_{\Psi_{\boldsymbol{\theta}}} - \beta^{-1}\langle \hat{S}_2 \rangle_{\Psi_{\boldsymbol{\theta}}} =   \dfrac{\bra{\Psi_{\boldsymbol{\theta}}}\hat{E}\ket{\Psi_{\boldsymbol{\theta}}}}{|\Psi_{\boldsymbol{\theta}}|^2} - \beta^{-1}\left( 1 -\dfrac{\bra{\Psi_{\boldsymbol{\theta}}}\hat{Q}\ket{\Psi_{\boldsymbol{\theta}}}}{|\Psi_{\boldsymbol{\theta}}|^4}\right)\,, 
    \label{eq_suppl_freeenergy}
\end{equation}
where $\hat{E} = \sum_{\mathbf{s}}E(\mathbf{s})\ketbra{\mathbf{s}}{\mathbf{s}}$, and $\hat{Q} =  \sum_{\mathbf{s}} |\braket{\mathbf{s}}{\Psi_{\boldsymbol{\theta}}}|^2 \ketbra{\mathbf{s}}{\mathbf{s}}$. Note that  $\langle \hat{E} \rangle_{\Psi_{\boldsymbol{\theta}}} \equiv \bra{\Psi_{\boldsymbol{\theta}}}\hat{E}\ket{\Psi_{\boldsymbol{\theta}}} / |\Psi_{\boldsymbol{\theta}}|^2$ and $\langle \hat{S}_2 \rangle_{\Psi_{\boldsymbol{\theta}}} \equiv 1 - \bra{\Psi_{\boldsymbol{\theta}}}\hat{Q}\ket{\Psi_{\boldsymbol{\theta}}} / |\Psi_{\boldsymbol{\theta}}|^4$ correspond to the average energy and the Tsallis entropy respectively. The gradient of the generalized free energy in Eq.~\eqref{eq_suppl_freeenergy} with respect to each of the MPS sites, $\left(\nabla_{\theta^{[n]}}\left[\cdot\right]\right)^{s}_{\alpha_{i}\alpha_{j}} \equiv \partial\left[\cdot\right]/\partial\theta^{[n]s}_{\alpha_{i}\alpha_{j}}$ , can be analytically computed as:  
\begin{equation}
    \nabla_{\boldsymbol{\theta}^{[n]}}\left[F\right] = \nabla_{\boldsymbol{\theta}^{[n]}}\left[\langle \hat{E} \rangle_{\Psi_{\boldsymbol{\theta}}} \right]+ \beta^{-1} \nabla_{\boldsymbol{\theta}^{[n]}} \left[\langle \hat{S}\rangle_{\Psi_{\boldsymbol{\theta}}}\right] \,,
\end{equation}
where 
\begin{equation}
    \nabla_{\boldsymbol{\theta}^{[n]}}\left[\langle \hat{E} \rangle_{\Psi_{\boldsymbol{\theta}}} \right] = \dfrac{\nabla_{\boldsymbol{\theta}^{[n]}} \left[\bra{\Psi_{\boldsymbol{\theta}}}\hat{E} \ket{\Psi_{\boldsymbol{\theta}}}\right] |\Psi_{\boldsymbol{\theta}}|^2 - \nabla_{\boldsymbol{\theta}^{[n]}} \left[|\Psi_{\boldsymbol{\theta}}|^2\right]\bra{\Psi_{\boldsymbol{\theta}}}\hat{E} \ket{\Psi_{\boldsymbol{\theta}}}}{|\Psi_{\boldsymbol{\theta}}|^4} \,,
\end{equation}
\begin{equation}
    \nabla_{\boldsymbol{\theta}^{[n]}}\left[\langle \hat{S} \rangle_{\Psi_{\boldsymbol{\theta}}}\right]  = -\dfrac{\nabla_{\boldsymbol{\theta}^{[n]}} \left[\bra{\Psi_{\boldsymbol{\theta}}}\hat{Q} \ket{\Psi_{\boldsymbol{\theta}}}\right] |\Psi_{\boldsymbol{\theta}}|^2 - 2 \nabla_{\boldsymbol{\theta}^{[n]}} \left[|\Psi_{\boldsymbol{\theta}}|^2\right]|\bra{\Psi_{\boldsymbol{\theta}}}\hat{Q} \ket{\Psi_{\boldsymbol{\theta}}}}{|\Psi_{\boldsymbol{\theta}}|^6} \,.
\end{equation}
Therefore, the generalized free energy optimization is reduced to the calculation, by means of efficient MPS contractions, of three observables: $\bra{\Psi_{\boldsymbol{\theta}}}\hat{E} \ket{\Psi_{\boldsymbol{\theta}}}$, $\bra{\Psi_{\boldsymbol{\theta}}}\hat{Q} \ket{\Psi_{\boldsymbol{\theta}}}$, and $|\Psi_{\boldsymbol{\theta}}|^2$; and their respective gradients: $\nabla_{\boldsymbol{\theta}^{[n]}} \left[\bra{\Psi_{\boldsymbol{\theta}}}\hat{E} \ket{\Psi_{\boldsymbol{\theta}}}\right]$, $\nabla_{\boldsymbol{\theta}^{[n]}} \left[\bra{\Psi_{\boldsymbol{\theta}}}\hat{Q} \ket{\Psi_{\boldsymbol{\theta}}}\right]$, and $\nabla_{\boldsymbol{\theta}^{[n]}} \left[|\Psi_{\boldsymbol{\theta}}|^2\right]$. The computational cost of the algorithm will then be largely determined by the Tensor Network contractions performed at each iteration to calculate the previous quantities. The order in which the indices of a Tensor Network are contracted affects the number of operations required for a computation, and thus influences the cost scaling of the algorithm. In this appendix, we present the cost of the optimal contraction strategy to compute all quantities involved in the algorithm. 

\begin{figure}
\includegraphics[width=1\linewidth]{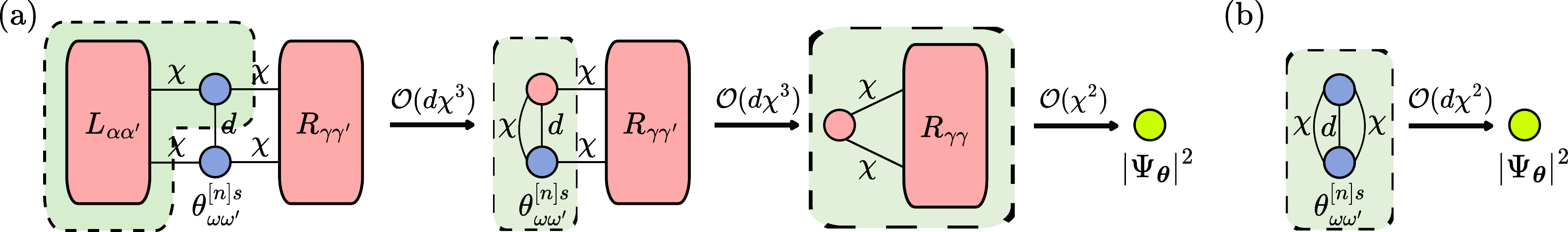}
\caption{Cost and order of contractions for the computation of the squared norm of an MPS, where $d$, $\chi$ denote the physical and bond dimension respectively, and $L_{\alpha\alpha'}$, $R_{\gamma\gamma'}$ are the left and right environments. We show that the cost for any MPS is $\mathcal{O}(\chi^3)$ in (a), and for the special case when MPS is in the canonical form is $\mathcal{O}(\chi^2)$ in (b) (Color online).}
\label{Fig:normcost}
\end{figure}

As diagrammatically shown in Fig.~\ref{Fig:normcost} (a), the optimal scheme to compute the squared norm $|\Psi_{\theta}|^2$ of a general MPS scales as $\mathcal{O}(d\chi^3)$, 
\begin{equation}
    |\Psi_{\theta}|^2 = \sum_{s=0}^{d-1}\sum_{\alpha\alpha'\gamma\gamma'=0}^{\chi} L_{\alpha\alpha'}\theta_{\alpha\gamma}^{[n]s}\theta_{\alpha'\gamma'}^{\dagger[n]s}R_{\gamma\gamma'} = \sum_{s=0}^{d-1}\sum_{\alpha'\gamma\gamma'=0}^{\chi} C_{\alpha'\gamma}^{s}\theta_{\alpha'\gamma'}^{\dagger[n]s}R_{\gamma\gamma'} = \sum_{\gamma\gamma'=0}^{\chi} C'_{\gamma'\gamma}R_{\gamma\gamma'} \,,
    \label{eqsupl_normcontraction}
\end{equation}
where $d$, $\chi$ denote the physical and bond dimension respectively, $L_{\alpha\alpha'}$, $R_{\gamma\gamma'}$ are the left and right environments, and $C$, $C'$ are auxiliary tensors. In the case of an MPS in the Canonical form, the environments reduce to the identity $L_{\alpha\alpha'}=\delta_{\alpha\alpha'}$, $R_{\gamma\gamma'}=\delta_{\gamma\gamma'}$, so that the squared norm calculation only needs one indices contraction with cost $\mathcal{O}(d\chi^2)$ as shown in Fig.~\ref{Fig:normcost} (b),
\begin{equation}
    |\Psi_{\theta}|^2 = \sum_{s=0}^{d-1}\sum_{\alpha\gamma=0}^{\chi} \theta_{\alpha\gamma}^{[n]s}\theta_{\alpha\gamma}^{\dagger[n]s} \,.
\end{equation}

\begin{figure}
\includegraphics[width=1\linewidth]{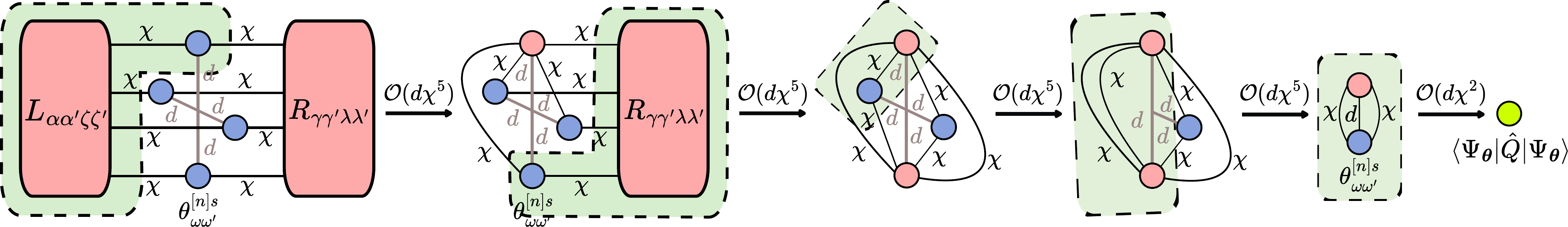}
\caption{Cost and order of contractions for the computation of the Tsallis entropy of an MPS, where $d$, $\chi$ denote the physical and bond dimension respectively, and $L_{\alpha\alpha'\zeta\zeta'}$, $R_{\gamma\gamma'\lambda\lambda'}$ are the left and right environments. We show that the cost is $\mathcal{O}(\chi^5)$. The semi-transparent labels $d$ denote the sum of the diagonal thus it scales as $\mathcal{O}(d)$ instead of $\mathcal{O}(d^4)$ (Color online).}
\label{Fig:entropycost}
\end{figure}
The non-normalized expectation value of the Hamiltonian is also computed at cost $\mathcal{O}(d\chi^3)$ by a similar scheme as the squared norm (Fig.~\ref{Fig:normcost}). The expected value of the Ising Hamiltonian can be decomposed as the sum of expectation values of two-local operators $\bra{\Psi_{\boldsymbol{\theta}}}\hat{E} \ket{\Psi_{\boldsymbol{\theta}}} = -\frac{1}{2}\sum_{i,j}J_{i,j}\bra{\Psi_{\boldsymbol{\theta}}}\hat{\sigma}^{[i]}\hat{\sigma}^{[j]}\ket{\Psi_{\boldsymbol{\theta}}}$ acting on MPS sites $i$ and $j$. Therefore, while the terms in which $i,j\neq n$ are computed as in Eq.~\eqref{eqsupl_normcontraction} with environments encompassing the action of the operators $\hat{\sigma}$, the elements where $i=n$ or $j=n$ are calculated as:
\begin{equation}
    \bra{\Psi_{\boldsymbol{\theta}}}\hat{\sigma}^{[i]}\hat{\sigma}^{[j]}\ket{\Psi_{\boldsymbol{\theta}}} = \sum_{ss'=0}^{d-1}\sum_{\alpha\alpha'\gamma\gamma'=0}^{\chi} L_{\alpha\alpha'}\theta_{\alpha\gamma}^{[n]s}\hat{\sigma}^{[n]s'}_{s}\theta_{\alpha'\gamma'}^{\dagger[n]s'}R_{\gamma\gamma'} = \sum_{s'=0}^{d-1}\sum_{\alpha'\gamma\gamma'=0}^{\chi} C_{\alpha'\gamma}^{s'}\theta_{\alpha'\gamma'}^{\dagger[n]s'}R_{\gamma\gamma'} = \sum_{\gamma\gamma'=0}^{\chi} C'_{\gamma'\gamma}R_{\gamma\gamma'} \,,
    \label{eqsupl_enercontraction}
\end{equation}
As for the the Tsallis entropy, Fig.~\ref{Fig:entropycost} shows the optimal strategy to compute $\bra{\Psi_{\boldsymbol{\theta}}}\hat{Q} \ket{\Psi_{\boldsymbol{\theta}}}$ at cost $\mathcal{O}(d\chi^5)$,
\begin{equation}
\begin{split}
    \bra{\Psi_{\boldsymbol{\theta}}}\hat{Q} \ket{\Psi_{\boldsymbol{\theta}}} & =  \sum_{s=0}^{d-1}\sum_{\alpha\alpha'\zeta\zeta'\gamma\gamma'\lambda\lambda''=0}^{\chi} L_{\alpha\alpha'\zeta\zeta'}\theta_{\alpha\gamma}^{[n]s}\theta_{\alpha'\gamma'}^{\dagger[n]s}\theta_{\zeta\lambda}^{[n]s}\theta_{\zeta'\lambda'}^{\dagger[n]s}R_{\gamma\gamma'\lambda\lambda'} = \sum_{s=0}^{d-1}\sum_{\alpha'\zeta\zeta'\gamma\gamma'\lambda\lambda'=0}^{\chi} C^s_{\gamma\alpha'\zeta\zeta'}\theta_{\alpha'\gamma'}^{\dagger[n]s}\theta_{\zeta\lambda}^{[n]s}\theta_{\zeta'\lambda'}^{\dagger[n]s}R_{\gamma\gamma'\lambda\lambda'} =
    \\ & = \sum_{s=0}^{d-1}\sum_{\alpha'\zeta\zeta'\gamma\gamma'\lambda=0}^{\chi} C^s_{\gamma\alpha'\zeta\zeta'}\theta_{\alpha'\gamma'}^{\dagger[n]s}\theta_{\zeta\lambda}^{[n]s}C^{'s}_{\gamma\gamma'\lambda\zeta'} = \sum_{s=0}^{d-1}\sum_{\zeta\zeta'\gamma\gamma'\lambda=0}^{\chi} C^{''s}_{\gamma\gamma'\zeta\zeta'}\theta_{\zeta\lambda}^{[n]s}C'_{\gamma\gamma'\lambda\zeta'} = \sum_{s=0}^{d-1}\sum_{\zeta\lambda=0}^{\chi} C^{'''s}_{\zeta\lambda}\theta_{\zeta\lambda}^{[n]s}
    \,.
\end{split}
\end{equation}

The corresponding gradients are exactly calculated at the same cost as the observables: 
\begin{equation}
    \nabla_{\boldsymbol{\theta}^{[n]}} \left[|\Psi_{\boldsymbol{\theta}}|^2\right]^s_{\alpha'\gamma'} = 2\sum_{\alpha\gamma=0}^{\chi} L_{\alpha\alpha'}\theta_{\alpha\gamma}^{[n]s} R_{\gamma\gamma'} = 2\sum_{\gamma=0}^{\chi} C_{\alpha'\gamma}^{s}R_{\gamma\gamma'} \,,  
\end{equation}
\begin{equation}
    \nabla_{\boldsymbol{\theta}^{[n]}} \left[\bra{\Psi_{\boldsymbol{\theta}}}\hat{\sigma}^{[i]}\hat{\sigma}^{[j]}\ket{\Psi_{\boldsymbol{\theta}}}\right]^s_{\alpha'\gamma'} = 2\sum_{s'=0}^{d-1} \sum_{\alpha\gamma=0}^{\chi} L_{\alpha\alpha'}\theta_{\alpha\gamma}^{[n]s'}\hat{\sigma}^{[n]s'}_{s} R_{\gamma\gamma'} = 2\sum_{\gamma=0}^{\chi} C_{\alpha'\gamma}^{s}R_{\gamma\gamma'}  \,, 
\end{equation}
\begin{equation}
\begin{split}
    \nabla_{\boldsymbol{\theta}^{[n]}} \left[\bra{\Psi_{\boldsymbol{\theta}}}\hat{Q} \ket{\Psi_{\boldsymbol{\theta}}}\right]^s_{\alpha'\gamma'} & = \sum_{\alpha\zeta\zeta'\gamma\lambda\lambda''=0}^{\chi} L_{\alpha\alpha'\zeta\zeta'}\theta_{\alpha\gamma}^{[n]s}\theta_{\zeta\lambda}^{[n]s}\theta_{\zeta'\lambda'}^{\dagger[n]s}R_{\gamma\gamma'\lambda\lambda'} = \sum_{\zeta\zeta'\gamma\lambda\lambda'=0}^{\chi} C^s_{\gamma\alpha'\zeta\zeta'}\theta_{\zeta\lambda}^{[n]s}\theta_{\zeta'\lambda'}^{\dagger[n]s}R_{\gamma\gamma'\lambda\lambda'} =
    \\ & = \sum_{\zeta\zeta'\gamma\lambda=0}^{\chi} C^s_{\gamma\alpha'\zeta\zeta'}\theta_{\zeta\lambda}^{[n]s}C^{'s}_{\gamma\gamma'\lambda\zeta'} =  \sum_{\zeta\lambda=0}^{\chi} C^{'''s}_{\zeta\lambda\alpha'\gamma'}\theta_{\zeta\lambda}^{[n]s}
    \,. 
\end{split}
\end{equation}
Diagrammatically, the calculation of the gradient can be represented as \textit{punching a hole} in the place of the MPS site $n$ (see Fig.~\ref{fig:tensor_network})~\cite{catarina2023density}.    

\begin{figure}
\includegraphics[width=0.5\linewidth]{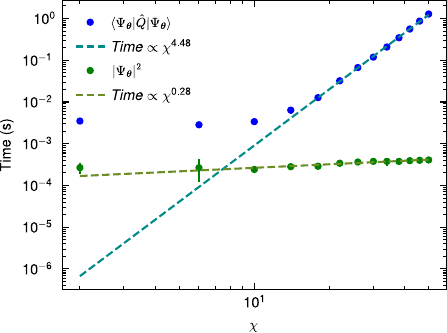}
\caption{Numerical scaling of the time with the MPS bond dimension $\chi$ to perform the tensor contractions involved by the observables computation and shown in Figs.~\ref{Fig:normcost} and~\ref{Fig:entropycost}. We observe how the numerical scaling is better than the analytical upper bound. The results show the average and standard deviation of 100 instances with random parameters performed by the \textit{opt-
einsum} Python library~\cite{g_a_smith_opt_einsum_2018} and the most optimal contraction scheme (Color online).}
\label{fig_numericalcostscaling}
\end{figure}

We numerically test the theoretical computational cost scalings derived above. For this purpose, we generate 100 instances of MPS sites, $\theta_{\alpha\gamma}^{[n]s}$, and left-right environments, $L_{\alpha\alpha'}$, $R_{\gamma\gamma'}$, $L_{\alpha\alpha'\zeta\zeta'}$, and $R_{\gamma\gamma'\lambda\lambda'}$. The instances are generated with random parameters and increasing bond dimension $\chi$ (100 samples for each bond dimension). For each instance, we compute the time it takes to calculate $|\Psi_{\boldsymbol{\theta}}|^2$ and $\bra{\Psi_{\boldsymbol{\theta}}}\hat{Q} \ket{\Psi_{\boldsymbol{\theta}}}$ using the Python library \textit{opt-
einsum}~\cite{g_a_smith_opt_einsum_2018} and the most optimal contraction scheme. The average and standard deviation of the results are shown in Fig.~\ref{fig_numericalcostscaling}). We observe that the theoretical cost indeed serves as an upper bound for the numerical scaling with the MPS bond dimension which uses efficient algorithms to perform specific contractions. 

\end{document}